\documentclass[11pt]{article}
\usepackage{amssymb, color}
\usepackage{bbm}
\usepackage{latexsym}
\usepackage{mathrsfs}
\usepackage{amsmath}
\usepackage{leqno}
\usepackage{amsfonts}
\usepackage{indentfirst}
\usepackage{graphics}
\usepackage[round]{natbib}
\usepackage{epsfig}
\usepackage[normalem]{ulem}
\usepackage{multirow}
\usepackage{slashbox}
\usepackage{longtable}
\usepackage{cite}

\usepackage{bm}
\def\bbeta{{\bm\beta}}
\def\bx{{\bm x}}
\def\btheta{{\bm\theta}}

\definecolor{dgreen}{rgb}{0,0.8,0}

\begin{document}
\label{werfefef}
\thispagestyle{empty}
\parindent 4mm
\baselineskip 20pt
\title
{\large\bf
    Estimation of Treatment Effects for Heterogeneous Matched Pairs Data with Probit Models}

\date{}

\author{\normalsize Jun Wang$^a$ \qquad Wei Gao$^{a\ *}$ \qquad Man-Lai Tang$^b$
\\[1ex]
\hspace{1em}
\normalsize
a Key Laboratory for Applied Statistics of MOE, School of Mathematics and Statistics, \\
\normalsize
Northeast Normal University, Changchun, Jilin 130024, China  \\
\normalsize
 b Department of Mathematics and Statistics, Hang Seng Management College, Hong Kong, China \\
}

\maketitle

\begin{abstract}
Estimating the effect of medical treatments on subject responses is one of the crucial problems in medical research. Matched-pairs designs are commonly implemented in the field of  medical research to eliminate confounding and improve efficiency. In this article, new estimators of treatment effects for heterogeneous matched pairs data are proposed. Asymptotic properties of the proposed estimators are derived. Simulation studies show that the proposed estimators have some advantages over the famous Heckman's estimator and inverse probability weighted (IPW) estimator. We apply the proposed methodologies to a blood lead level data set and an acute leukaemia data set.
\end{abstract}


\section{Introduction}
In many medical studies, researchers are often interested in evaluating the treatment effects for some improved surgery or new medicine, and one popular measurement is average treatment effects (ATE)
$$
ATE=E[Y(1)-Y(0)]
$$
 where $Y(0)$ and $Y(1)$ denote the two potential outcomes given the control treatment ($D=0$) and the active treatment($D=1$). Neyman (1923) originally proposed potential outcomes models, Rubin (1974) generalized it, which is usually called Rubin causal models(RCM). In order to effectively estimate treatment effects, one challenge problem is to control confounding, and Abadie and Imbens (2006, 2011), D\'{i}az, {\it et al.} (2015) proposed covariates balanced methods to estimate average treatment effects (ATE) under strong ignorability conditions.
\par
 Another method to control confounding is the matched pairs studies (e.g., twins study), in which one of the subjects is randomly selected to expose to treatments and the other is selected to receive the control treatment. Let $A$ and $B$ be the pairs with covariates $\bx^A$ and $\bx^B$ respectively, $D$ denote the treatment  status with $D=0$ for control and $D=1$ for new treatment, and $Y^A$ be the response value for subject $A$ with treatment status $D$ and $Y^B$ is the response value for subject $B$ with treatment status $1 - D$. When response value is linearly relationship with treatment status, treatment effects are usually estimated by difference-in-differences (DID) approaches (Ashenfelter and Card, 1985; Imbens and Angrist, 1994; Card and Krueger, 1994; Abadie and Gardeazabal, 2003). Lately, Sakaguchi (2016) considered the time-varying treatment effects model for panel data and estimated the time-varying average treatment effects by difference-in-differences (DID) approaches.
\par
However, in practice, in terms of matched pairs data, the relationship between response value and treatment status is usually nonlinear, namely, the response value is assumed to be  binary with $0$ representing negative effect and $1$ representing positive effect. The following model is assumed:
\begin{equation}
Y^A=I({\bbeta}^{'}{\bx}^{A}+\lambda D+\eta^A>0), \mbox{~and}\;\;\;Y^B=I(\bbeta^{'}\bx^{B}+\lambda (1-D)+\eta^B>0) \label{MI}
\end{equation}
where $\bbeta$ and $\lambda$ are $k\times 1$ regression coefficients and treatment parameters respectively, and $\eta^A$ and $\eta^B$ are errors.
\par
If $\eta^A$ and $\eta^B$ in Model (\ref{MI}) are independent of $\bx^A$, $\bx^B$ and $D$, various authors have considered the problem of estimating the treatment effect. For example, Butler and Moffitt (1982) proposed a computationally efficient quadrature procedure to obtain the ML (maximum likelihood) estimators for the one-factor multinomial probit model while Bertschek and Lechner (1998) proposed a set of convenient GMM (generalized methods of moments) estimators for the binomial probit model based on panel data. Rubin (1974) presented a discussion of matching, randomization, random sampling, and other methods of controlling extraneous variation. Robins, {\it et al.} (1994), Robins and Rotnizky (1995), Hirano, {\it et al.} (2003), Hernan and Robins (2006) proposed the inverse probability weighted estimator for treatment effects.
\par
In practice, matched pairs data are collected from different backgrounds and it is not uncommon that some variables may not be observed for some pairs. Besides, although paired observations may share common features (e.g., same genes or environments), there may exist a great deal of heterogeneity among paired observations. For non-linear models, estimation of treatment effects in the presence of heterogeneity is a challenging problem. Heckman {\it et al.} (1997) gave detailed discussions on this issue and proposed the following models
\begin{equation}
Y^A=I({\bbeta}^{'}{\bx}^{A}+\lambda D+\tau+\epsilon^A>0), \mbox{~and}\;\;\;Y^B=I(\bbeta^{'}\bx^{B}+\lambda (1-D)+\tau+\epsilon^B>0) \label{MB}
\end{equation}
where $\bbeta$ and $\lambda$ are $k\times 1$ regression coefficients and treatment parameters respectively, $\tau$ is the unobserved grouped effects (e.g., twins share common genes) and is independent of $\epsilon^A$ and $\epsilon^B$, and $\epsilon^A$ and $\epsilon^B$ are errors with means $0$ and variances $1$.
\par
It is noteworthy that $\tau$ given in Models $(\ref{MB})$ can be considered a constant (i.e., fixed effect model) or random variable (i.e.,  random effect models). If we assume that $\tau$ is a random variable, there's usually have a prior information about group effects. Hence, assuming that $\epsilon^A$ and $\epsilon^B$ are normally distributed and $\tau$ has prior normal distributions, Heckman (1978) developed the maximum likelihood and other estimates based on the multivariate probit model with structural shift  to estimate $\bbeta$ and $\lambda$. However, Heckman's estimates can be very sensitive to the  prior distribution for $\tau$. Estimation problems of random effect models is discussed in Guilkey and Murphy(1993), Arellano and Bonhomme (2009), Walker (1996), Dey {\it et al.} (1997), Horowitz (1992) and Arulampalam and Stewart (2009). Recently Gao {\it et al.} (2017) proposed a new method to estimate the parameters for dynamic probit models.

\par
When $\tau$ is fixed effect, different estimation methods have been considered by various authors for panel data. For instance, Arellano (2003) applied the orthogonal technique to analyze discrete models for panel data. Carro (2007) demonstrated that the orthogonal technique can reduce the order bias of the maximum likelihood estimator from $O(T^{-1})$ to $O(T^{-2})$  where $T$ is the length of individual observations. Bartolucci {\it et al.} (2016) proposed the modified profile likelihood estimator for fixed-effects panel data models.

\par
In this article, we will consider models (\ref{MB}) when $\epsilon^A$ and $\epsilon^B$ are normally distributed with means $0$ and variances $1$ (i.e., probit model). Estimators for $\bbeta$ and $\lambda$ are considered and their theoretical properties  are presented in Section 2. Simulation studies will be conducted and the results will be reported in Section 3. In Section 4, we demonstrate our methodologies with two real data sets. Conclusions will be discussed in Section 5. Proofs of lemmas and theorems will be presented in Appendix.

\section{Probit Model and Estimator }
By assuming that $\epsilon^A$ and $\epsilon^B$ are normally distributed with means being $0$ and variances being $1$, Models (\ref{MB}) becomes the famous Probit Models. We aim to estimate $\lambda$ and $\bbeta$ of model (\ref{MB}) when $D$ is randomly selected between $0$ and $1$.
 \par
 Let $(D_1,Y^{A}_1,Y^{B}_1,{\bx}^{A}_1,{\bx}^{B}_1),\cdots,(D_n,Y^{A}_n,Y^{B}_n,{\bx}^{A}_n,{\bx}^{B}_n)$ be $n$ observations with
\begin{equation}
Y^A_i=I(\bbeta^{'}{\bx}_i^{A}+\tau_i+\lambda D_i+\epsilon^{A}_i>0), \mbox{~and}\;\;Y_i^B=I(\bbeta^{'}{\bx}_i^{B}+\tau_i+\lambda (1-D_i)+\epsilon^{B}_i>0)\label{ME}
\end{equation}
where $\bbeta$ and $\lambda$ are $k \times 1$ regression coefficients and treatment parameters respectively, $\tau_i$ is unobserved grouped effects and independent of $\epsilon^A_i$ and $\epsilon^B_i$, and $\epsilon^A_i$ and $\epsilon^B_i$ are independently normally distributed with means $0$ and variances $1$.
\par
When there are no covariates, estimating the treatment effects of $D$ is of the major interest and one usually considers the average treatment effect (ATE), i.e.,
\begin{equation*}
ATE=E[Y^A D+Y^B(1-D)-(Y^A(1-D)+Y^BD)] .
\end{equation*}
The average treatment effect (ATE) can be simply estimated by the sample $\{Y_i^A, {\bf x}^A_i, D_i\}_{i=1}^{n}$, $\{Y_i^B, {\bf x}^B_i, 1-D_i\}_{i=1}^{n}$, i.e.,
 \begin{equation*}
  \widehat{ATE}=\frac{1}{n}\sum \limits_{i}^{n} \Bigg(Y_i^A D_i+Y_i^B(1-D_i)-(Y_i^A(1-D_i)+Y_i^BD_i) \Bigg).
  \end{equation*}
 When the response is a linear function of $D$ or individuals are homogeneous, the average treatment effect is an efficient estimator for measuring the treatment effect of $D$. Under the Probit model (\ref{ME}), it is noteworthy that the treatment effect is unidentifiable if individuals demonstrate substantial heterogeneity. For example,
\begin{eqnarray*}
ATE &=& E[Y^A D+Y^B(1-D)-(Y^A(1-D)+Y^BD)]\\
&=&E[Y^A\mid D=1]p(D=1)+E[Y^B \mid D=0]p(D=0)\\
&&-E[Y^B\mid D=1]p(D=1)-E[Y^A \mid D=0]p(D=0) \\
&=&\int \Phi(\lambda+\tau) f(\tau)d\tau-\int \Phi(\tau) f(\tau)d\tau
=\Phi\left(\frac{\lambda-\mu_{\tau}}{\sqrt{1+\sigma_{\tau}^2}}\right)-\Phi\left(\frac{-\mu_{\tau}}{\sqrt{1+\sigma_{\tau}^2}}\right)
\end{eqnarray*}
when $\tau$ is normally distributed with mean $\mu_\tau$ and standard deviation $\sigma_{\tau}$, or
\begin{eqnarray*}
ATE &=& E[Y^A D+Y^B(1-D)-(Y^A(1-D)+Y^BD)]\\
&=&E[Y^A\mid D=1]p(D=1)+E[Y^B \mid D=0]p(D=0)\\
&&-E[Y^B\mid D=1]p(D=1)-E[Y^A \mid D=0]p(D=0) \\
&=&\int \Phi(\lambda+\tau) f(\tau)d\tau-\int \Phi(\tau) f(\tau)d\tau   \\
&=&p \left[\Phi\left(\frac{\lambda-\mu_{\tau_1}}{\sqrt{1+\sigma_{\tau_1}^2}}\right)-\Phi\left(\frac{-\mu_{\tau_1}}{\sqrt{1+\sigma_{\tau_1}^2}}\right) \right]+
(1-p)\left[\Phi\left(\frac{\lambda-\mu_{\tau_2}}{\sqrt{1+\sigma_{\tau_2}^2}}\right)-\Phi\left(\frac{-\mu_{\tau_2}}{\sqrt{1+\sigma_{\tau_2}^2}}\right)\right]
\end{eqnarray*}
when $\tau$ is a mixture of two normal variates, e.g., $p*N(\mu_1,\sigma_{\tau_1}^2)+(1-p)*N(\mu_2,\sigma_{\tau_2}^2)$. The average treatment effect is close to zero and unidentifiable if the mean or variance of the group effect $\tau$ is large.
\par
 In practical, collected data specially for observational studies exist heterogeneous among groups. Following Gao {\it et al.} (2017), we propose new estimators for the treatment effect $\lambda$ and $\bbeta$ in model (\ref{ME}) which can overcome the unidentifiable problem when the variance of individual effects (i.e., $\tau$) is large. Under the condition that $D_i$ is randomly selected between $0$ and $1$ and is independent of individual effect $\tau_i$, the following estimation procedure for $\lambda$ and $\bbeta$ is proposed
\begin{equation}
(\hat{\bbeta},\hat{\lambda})=\arg\max\prod_{i=1}^{n}\left\{ p_i^{Y^A_i(1-Y_i^B)}[1-p_i]^{(1-Y^A_i)Y_i^B}\right\}\label{MF},
\end{equation}
where
$$
p_i=p({\bbeta,\lambda};\bx_i^{A},\bx_i^{B},D_i)=\frac{G(\lambda(1-2D_i)+\bbeta^{'}(\bx^B_i-\bx^A_i))}{G(\lambda(1-2D_i)+\bbeta^{'}(\bx^B_i-\bx^A_i))+G(-\lambda(1-2D_i)-\bbeta^{'}(\bx^B_i-\bx^A_i))} $$
and
$$
G(x)=-\sqrt{\pi}x\Phi(-\frac{x}{\sqrt{2}})+\exp\{-\frac{x^2}{4}\}.
$$
\par
Let
\begin{equation}
L(\bbeta,\lambda)=\prod_{i=1}^{n}\left\{ p_i^{Y^A_i(1-Y_i^B)}[1-p_i]^{(1-Y^A_i)Y_i^B}\right\}\label{LF}.
\end{equation}
Here, (\ref{LF}) can be regarded as a conditional likelihood functions with $Y^A_i+Y^B_i=1$ while (\ref{MF}) can be viewed as a conditional maximum likelihood estimation problem. The identifiable problem for (\ref{MF}) is reported in the following theorem and its proof can be obtained by similar method given by Gao {\it et al.} (2017).
\par

{\bf Theorem 1.} There exists an unique solution (i.e., $(\hat{\bbeta}, \hat{\lambda}$))  to (\ref{MF}) if the following conditions hold
\begin{description}
\item (a) the rank of
 $$
    \left\{(\bx^B_{1}-\bx^A_{1})I(Y^A_1+Y^B_1=1),\cdots,(\bx^B_{n}-\bx^A_{n})I(Y^A_n+Y^B_n=1)\right\}
    $$
     is $k$ (i.e., the number of covariates), and
 \item (b) there exists some $j$ such that $D_j=0$ and
    $
    \bx^B_{j}-\bx^A_{j}=\sum\limits_{i\not=j}d_i(\bx^B_{i}-\bx^A_{i})I(D_i=0)
    $
    where $d_1,\cdots,d_{j-1},d_{j+1},\cdots,d_n$ are non-positive numbers.
\end{description}

\par
If  individuals $A$ and $B$ are sampled from the same population,  condition of (b) given in Theorem 1 is satisfied with probability close to $1$ for large $n$. Condition (b) given in Theorem 1 can also be rewritten as
 \begin{description}
 \item (b$^{'}$) there exists some $j$ such that $D_j=1$ and
    $
    \bx^B_{j}-\bx^A_{j}=\sum\limits_{i\not=j}d_i(\bx^B_{i}-\bx^A_{i})I(D_i=1)
    $
    where $d_1,\cdots,d_{j-1},d_{j+1},\cdots,d_n$ are non-positive numbers.
 \end{description}

\par
Gao {\it et al} (2017) developed estimators for the dynamic parameters in dynamic models and they are  consistent and asymptotically normal distributed. We will show that $\hat{\bbeta}$ and $\hat{\lambda}$ are also consistent and asymptotically normal distributed under some regular conditions.  For the sake of simplicity, denote $\hat{\btheta}=(\hat{\bbeta}^{'},\hat{\lambda})^{'}$ and $\btheta=(\bbeta^{'},\lambda)^{'}$.

\par
{\bf Assumption:} The individual effect is continuously distributed with the density function $f(x)$ which satisfies
\begin{equation}
f(x)=\frac{1}{\sigma_{\tau}}h\left(\dfrac{x-\mu_{\tau}}{\sigma_{\tau}}\right).\label{AS1}
\end{equation}

Under Assumption (\ref{AS1}), $\hat{\btheta}$ is consistent and asymptotically normally distributed and its proof is given in Appendix A.

\par
{\bf Theorem 2.} (Consistency) Under Assumption (\ref{AS1}) with $\sigma_{\tau}=an^{\alpha}$ $(0<\alpha<1)$ for some constant $a>0$,
$\hat{\btheta}$ converges to $\btheta$ in probability $1$ as $n\rightarrow\infty$.

\par
Let
\begin{eqnarray*}
K(\btheta;\bx^A,\bx^B,D)&=&\int \Phi(\bbeta^{'}\bx^A+\lambda D+u) \Phi(-\bbeta^{'}\bx^{B}-\lambda(1-D)-u)du\\
&&+
\int \Phi(-\bbeta^{'} x^{A}-\lambda D-u) \Phi(\bbeta^{'}x^{B}+\lambda(1-D) +u)du.
\end{eqnarray*}
Noting that
$$
p({\btheta};\bx^{A}_i,\bx^{B}_i,D_i)=\frac{\int \Phi(\bbeta^{'}\bx^A_i+\lambda D_i+u) \Phi(-\bbeta^{'}\bx^{B}_i-\lambda(1-D_i)-u)du}
{K(\btheta;\bx^A_i,\bx^B_i,D_i)}
$$
and
$$
1-p({\btheta};\bx^{A}_i,\bx^{B}_i,D_i)=\frac{\int \Phi(-\bbeta^{'}\bx^A_i-\lambda D_i-u) \Phi(\bbeta^{'}\bx^{B}_i+\lambda(1-D_i)+u)du}
{K(\btheta;\bx^A_i,\bx^B_i,D_i)},
$$
\noindent we have the following result
\par

{\bf Theorem 3.} (Asymptotic Normality) Let $h(x)$ be continuously derivative and  $\sigma_{\tau}=a\sqrt{n}$($a>0$). Suppose that the covariance matrix of $\bx^A-\bx^B$ is positive definite, the conditional distribution of $\bx^A-\bx^B$ given $D=0$ is symmetrical at the origin and $0<P[D=1\mid x]<1$. Under Assumption (\ref{AS1}) , we have
$k_{n}(\hat{\btheta}-{\btheta})$ is asymptotically distributed as $N(0,c\Sigma^{-1})$ as $n\rightarrow\infty$,
where
$$
k_{n}=\left\{\sum\limits_{i=1}^{n}I {(Y^{A}_{i}+Y^{B}_{i}=1)}\right\}^{1/2}
$$
and
$$
\Sigma=E\left\{
\frac{K(\btheta;\bx^A,\bx^B,D)\Big(\frac{\partial p({\btheta}; \bx^{A},\bx^{B},D)}{\partial {\btheta}}\Big)\Big(\frac{\partial p({\btheta};\bx^A,\bx^B,D)} {\partial {\btheta}\partial }\Big)^{'}}{p({\btheta}; \bx^{A},\bx^{B},D)\times(1-p({\btheta};\bx^{A},\bx^{B},D))}\right\},\mbox{~and}\;\;c=E[K(\btheta;\bx^A,\bx^B,D)].
$$

\section{Simulations}
In this section, we evaluate the finite sample performances of the treatment effects estimator $\hat{\lambda}$ which is proposed in Section 2. Several scenarios will be considered.

\par
In the first simulation study, we mainly consider models $(\ref{ME})$ without covariates. In this case, the error terms, i.e., $\epsilon_i^A$ and $\epsilon_i^B$, are assumed to be i.i.d. normally distributed with means being 0 and variances being 1 and $D_i$ follows binomial distribution with success probability being $p=2/3$. We assume that the individual effects come from different distributions (e.g., uniform, normal, Student t and Cauchy distributions). We obtain the mean bias (BIAS) and root mean squared errors (RMSEs) of the estimates of different $\lambda$ for different distributions of $\tau$ with sample sizes being 1000 or 5000. The results based on 100 repetitions are reported in Table $\ref{DA}$. From Table $\ref{DA}$, we observe that our proposed estimator is robust to different prior distributions for $\tau$. In particular, it is noteworthy that our proposed estimator works well for long-tailed distribution (i.e., Student-t) and distribution with no moments (i.e., t(1)(Cauchy)).

\begin{table}
\newcommand{\tabincell}[2]{\begin{tabular}{@{}#1@{}}#2\end{tabular}}
\renewcommand{\tabcolsep}{0.4pc}
\renewcommand{\arraystretch}{0.6}
  \centering
   \caption{Simulated RMSEs and BIAS of the new estimator for  treatment effect $\lambda $ in model $(\ref{ME})$ without covariates based on 100 replications.}
  \begin{tabular}{c|ccccccc}\hline
  \backslashbox{sample size}{ $\tau_i$} &parameter  &U(-4,4)& &N(0,4)&&t(1)&  \\ \hline
  & $\lambda $& BIAS &RMSE&BIAS & RMSE& BIAS &  RMSE   \\  \hline

n=1000 &  \tabincell{c}{2\\1.5\\1\\0.5\\0\\-0.5\\-1\\-1.5\\-2\\ } & \tabincell{c}{0.022\\0.011\\0.018\\0.002\\-0.018\\-0.006\\-0.053\\-0.030\\-0.026\\  } &\tabincell{c}{ 0.184\\0.150\\0.126\\0.103\\0.099\\0.091\\0.122\\0.158\\0.242\\ } & \tabincell{c}{0.019\\0.000\\-0.016\\-0.010\\-0.011\\0.004\\-0.019\\-0.003\\-0.053\\} &\tabincell{c}{0.256\\0.178\\0.124\\0.126\\0.114\\0.107\\0.144\\0.171\\0.254\\} &\tabincell{c}{0.026\\-0.013\\-0.046\\-0.012\\-0.008\\0.022\\0.038\\0.025\\-0.025\\} &\tabincell{c}{0.236\\0.143\\0.109\\0.093\\0.092\\0.082\\0.109\\0.128\\0.206\\} \\\hline

\backslashbox{sample size}{ $\tau_i$}&$\lambda$  &U(-10,10)& &N(0,25)&&t(1)&  \\\hline

  n=5000 & \tabincell{c}{2\\1.5\\1\\0.5\\0\\-0.5\\-1\\-1.5\\-2\\ }  & \tabincell{c}{0.038\\-0.005\\0.006\\0.001\\0.003\\0.013\\-0.002\\-0.006\\-0.030\\} &\tabincell{c}{0.180\\0.114\\0.079\\0.062\\0.062\\0.071\\0.071\\0.097\\0.151\\}  & \tabincell{c}{0.039\\0.055\\0.028\\0.024\\-0.011\\0.014\\-0.015\\-0.011\\-0.056\\} &\tabincell{c}{0.283\\0.202\\0.160\\0.140\\0.123\\0.115\\0.136\\0.228\\0.271\\}&\tabincell{c}{-0.008\\-0.008\\-0.035\\-0.005\\0.003\\0.018\\0.029\\0.014\\0.011\\}& \tabincell{c}{0.075\\0.066\\0.054\\0.037\\0.033\\0.037\\0.054\\0.038\\0.073\\}  \\\hline
\end{tabular}
\label{DA}
\end{table}
\par
Next, we consider models $(\ref{ME})$ without covariates and with the individual effect $\tau_i \sim N(0, \pi^2/3)$ where $\pi$ = 3.14. The results for standard error (SE), BIAS and RMSE of our proposed  estimate for various $\lambda$-values based on sample sizes being 200, 500, 1000, 2000, 3000 and 5000 are reported in Table $\ref{DB}$. We find that as the sample size increases,  all BIAS, RMSE and SE for $\hat{\lambda}$ decrease, which is consistent with Theorem 2 that $\hat{\lambda}$ is a consistent estimator of $\lambda$. Besides, we observe that both SE and RMSE are almost identical. In addition, we consider model $(\ref{ME})$ with covariates in which $ x_{i}^{A}$ being sampled from N(0, 1) and $x_{i}^{B} = x_{i}^{A} + N(0,1)$. Mean BIASs and RMSEs of the estimates of $\lambda$ and $\beta$ based on 100 repetitions for different values of $\lambda$ and $\beta$ with sample size being 1000 are reported in Table $\ref{DC}$. We find that the biases for both $\hat{\lambda}$ and $\hat{\beta}$ are close to zero, which again confirms the result of Theorem 2 that $\hat{\btheta}$ is a consistent estimator of ${\btheta}$.
\par
\begin{table}
\newcommand{\tabincell}[2]{\begin{tabular}{@{}#1@{}}#2\end{tabular}}
  \centering
   \caption{Simulated RMSEs, BIAS and SE of the new estimator  for  treatment effect $\lambda $  in models $(\ref{ME})$ without covariates and $\tau_i \ \sim \  N(0,\pi^2/3)$ based on 100 replications.}
  \begin{tabular}{ccccc|ccccc}\hline
   sample size  & $\lambda$ & BIAS  & SE & RSME &  sample size & $\lambda$ & BIAS  & SE &RMSE  \\ \hline
\tabincell{c} { \\  \\200  \\  \\  \\} & \tabincell{c}{1 \\ 0.5 \\0\\ -0.5 \\ -1 \\} &\tabincell{c} {0.015 \\0.003\\  0.003\\-0.029\\-0.029 \\}&\tabincell{c}{0.295\\0.259\\0.236\\0.248\\0.293 } &\tabincell{c}{0.294  \\ 0.258\\ 0.235\\0.248\\ 0.293\\} & 2000 &\tabincell{c}{1 \\ 0.5 \\0\\ -0.5 \\ -1 \\}  &\tabincell{c}{-0.007 \\ 0.004\\-0.006\\0.009\\0.011\\}& \tabincell{c}{0.088\\0.068 \\0.069\\0.076\\0.087\\} & \tabincell{c}{0.088\\0.068 \\ 0.069\\0.076\\0.087\\} \\\hline

   500& \tabincell{c} {1 \\ 0.5 \\0\\ -0.5 \\ 1 \\} &\tabincell{c} {0.016 \\0.001\\ -0.018 \\0.001\\-0.029\\}&\tabincell{c}{0.164\\0.133\\0.127\\0.143\\0.177 } &\tabincell{c}{0.164 \\0.132\\ 0.128\\0.143\\0.178\\ } & 3000 & \tabincell{c}{1 \\ 0.5 \\0\\ -0.5 \\ -1 \\} & \tabincell{c}{  -0.008\\ 0.002\\ -0.004\\-0.001\\0.009 \\ } & \tabincell{c}{0.070\\ 0.059 \\-0.004\\0.062 \\0.077 \\} & \tabincell{c}{0.070 \\ 0.059  \\ 0.059 \\  0.062 \\0.077\\} \\\hline

     1000 & \tabincell{c} {1 \\ 0.5 \\ 0\\-0.5 \\ -1 \\} &\tabincell{c} {-0.005\\-0.016\\ -0.010\\0.003\\0.001\\}&\tabincell{c}{0.116\\0.101\\0.093\\0.105\\0.126} &\tabincell{c}{0.115\\0.102\\0.093\\0.105\\0.125 }&5000 &  \tabincell{c}{ 1\\0.5\\0\\-0.5\\-1\\}  &\tabincell{c}{ -0.006\\-0.001\\ 0.004\\-0.001\\0.004\\} & \tabincell{c}{0.054\\0.041\\0.042\\0.047\\0.053\\} & \tabincell{c}{  0.054\\0.041\\0.042\\0.047\\0.053\\} \\\hline
\end{tabular}
\label{DB}
\end{table}

\begin{table}
 \newcommand{\tabincell}[2]{\begin{tabular}{@{}#1@{}}#2\end{tabular}}
  \renewcommand{\tabcolsep}{0.01pc}
\renewcommand{\arraystretch}{1.0}
  \centering
   \caption{Simulated RMSEs and BIAS of the new estimator of the  parameter in models $(\ref{ME})$ with covariates, sample size n=1000, $\tau_i \ \sim \ N(0,\pi^2/3)$ based on 100 replications.}
  \begin{tabular}{cccccc|cccccc}\hline
   $\lambda$  & $\bbeta$ & BIAS$(\hat{\lambda})$  & RMSE$(\hat{\lambda})$ &BIAS$(\hat{\bbeta})$ & RMSE$(\hat{\bbeta})$ & $\lambda$  & $\bbeta$ &BIAS$(\hat{\lambda})$& RMSE$(\hat{\lambda})$ &BIAS$(\hat{\bbeta})$  & RMSE$(\hat{\bbeta})$ \\ \hline
\tabincell{c}{1 \\ 0.5 \\0\\ -0.5 \\ -1 \\}  & \tabincell{c}{1 \\ 0.5 \\0\\ -0.5 \\ -1 \\} &\tabincell{c} {0.029\\0.011\\-0.001\\-0.016\\-0.015\\}  &\tabincell{c} {0.029\\0.011\\-0.001\\-0.016\\-0.015\\} &\tabincell{c} {0.025\\-0.011\\0.041\\-0.025\\-0.027\\} & \tabincell{c}{0.170\\0.108\\0.105\\0.109\\0.179\\} & \tabincell{c}{1 \\ 0.5 \\ \\ -0.5 \\ -1 \\}  &  \tabincell{c}{-1 \\- 0.5 \\ \\ 0.5 \\ 1 \\}&\tabincell{c}{0.000\\-0.017\\ \\0.010\\-0.015\\}&\tabincell{c}{0.177\\0.118\\ \\0.117\\0.193 \\} &\tabincell{c}{-0.019\\0.001\\ \\0.011\\0.042\\}&\tabincell{c}{0.172\\0.095\\  \\ 0.093\\0.186\\} \\
\\
   \tabincell{c}{0  \\0\\ 0 \\ 0 \\} & \tabincell{c} {1 \\ 0.5\\-0.5\\ -1 \\} &\tabincell{c}{0.005\\0.088\\ 0.015\\-0.024\\}&\tabincell{c} {0.133\\0.103\\0.102\\0.138\\} &\tabincell{c}{0.016\\0.020\\ 0.001\\-0.014\\} & \tabincell{c}{0.165\\0.088\\0.094\\0.145\\} &  \tabincell{c}{1 \\ 0.5 \\ -0.5 \\ -1 \\} & \tabincell{c}{0\\0\\0\\0\\} &\tabincell{c}{0.024\\0.016\\ 0.001\\-0.005\\}&\tabincell{c}{0.145\\0.109\\0.105\\0.126\\}& \tabincell{c}{ 0.016\\0.008\\-0.007\\-0.014\\ }&  \tabincell{c}{0.097\\0.081\\0.078\\0.094\\} \\\hline

\end{tabular}
\label{DC}
\end{table}
\par
For normally distributed group effects with mean being 0 and variance being $\sigma^2$ in models $(\ref{ME})$, Heckman (1978) proposed the maximum likelihood estimators for the parameter $\lambda$ and $\sigma^2$ (denoted as $\hat{\lambda}_H$  and $\hat{\sigma}_H$, respectively). Here, we compare our proposed estimators with Heckman estimators for the random effect in models $(\ref{ME})$ without covariates. For this purpose, we consider the group effects coming from N(0,1), N(0,4), and symmetrical mixed normal distribution $0.5*N(-6,9)+0.5*N(6,9)$. BIASs and RMSEs of the estimates (based on 100 repetitions) for different $\lambda$ values with sample size being 1000 are reported  in Table $\ref{DD}$. We also consider group effect coming from asymmetric mixed normal distribution $0.5*N(-6,9)+0.5*N(6,3)$, $0.5*N(-6,3)+0.5*N(6,9)$ and the results are reported in Table $\ref{DD}$. From Table $\ref{DD}$, we find that our proposed estimator and Heckmans estimator are comparable. However, for asymmetrically distributed group effects, our proposed estimator is significantly better that Heckmans estimator in terms of smaller BIASs and RMSEs.
\begin{table}
\newcommand{\tabincell}[2]{\begin{tabular}{@{}#1@{}}#2\end{tabular}}
  \renewcommand{\tabcolsep}{0.05pc}
\renewcommand{\arraystretch}{0.4}
  \centering
   \caption{Comparison of RMSEs and BIAS of our proposed estimator $\hat{\lambda}_{new} $ and Heckmans estimator $\hat{\lambda}_{H}$ for various group effect distribution with sample size being 1000 based on 100 repetitions.}
  \begin{tabular}{cccccccc}\hline
 Distribution of the $\tau_i$  & $\lambda$ &BIAS$(\hat{\lambda}_{new})$& RMSE$(\hat{\lambda}_{new})$  &BIAS$(\hat{\lambda}_H)$ & RMSE$(\hat{\lambda}_H)$ &BIAS$(\hat{\sigma}_H)$ & RMSE$(\hat{\sigma}_H)$  \\ \hline
N(0,1) & \tabincell{c}{1 \\ 0.5 \\0\\ -0.5 \\ -1 \\} &\tabincell{c} {-0.027\\-0.025\\0.000\\0.021\\0.031\\} &\tabincell{c} {0.092\\0.061\\0.059\\0.083\\0.088\\}&\tabincell{c} {0.018\\0.008\\0.003\\0.004\\0.002\\} & \tabincell{c}{0.077\\0.064\\0.051\\0.058\\0.070\\} &\tabincell{c} {0.011\\0.009\\0.005\\0.002\\0.002\\} & \tabincell{c}{0.084\\0.084\\0.082\\0.082\\0.095\\}   \\\hline
   N(0,4) & \tabincell{c} {1 \\ 0.5\\0\\-0.5\\ -1 \\} &\tabincell{c} {-0.016\\0.010\\-0.011\\0.004\\-0.019} &\tabincell{c} {0.124\\0.126\\0.114\\0.107\\0.144\\}&\tabincell{c}{-0.005\\0.010\\-0.008\\0.018\\-0.011\\}&\tabincell{c}{0.138\\0.121\\0.121\\0.105\\0.138\\} &\tabincell{c}{0.057\\0.072\\0.062\\0.078\\0.089\\}&\tabincell{c}{0.470\\0.472\\0.337\\0.431\\0.471\\}   \\\hline

 \tabincell{c} {0.5*N(-6,9) \\+0.5*N(6,9)\\} & \tabincell{c} {1 \\ 0.5\\0\\-0.5\\ -1 \\}& \tabincell{c} {0.043\\0.014\\0.002\\-0.014\\-0.014\\}&\tabincell{c} {0.274\\0.174\\0.186\\0.200\\0.251\\}&\tabincell{c} {0.036\\-0.029\\0.010\\-0.015\\-0.093\\}&\tabincell{c}{0.251\\0.206\\0.185\\0.202\\0.285 \\} &\tabincell{c} {1.033\\1.083\\0.756\\0.943\\1.173\\}&\tabincell{c}{2.462\\2.426\\2.110\\2.034\\2.850\\}   \\\hline
  \tabincell{c} {0.5*N(-6,9) \\+0.5*N(6,3)\\} & \tabincell{c} {1 \\ 0.5\\0\\-0.5\\ -1 \\}& \tabincell{c} {0.052\\0.011\\-0.001\\-0.010\\-0.051\\}&\tabincell{c} {0.303\\0.221\\0.206\\0.211\\0.314\\}&\tabincell{c} {0.327\\0.303\\0.273\\0.232\\0.201\\}&\tabincell{c} {0.427\\0.387\\0.340\\0.251\\0.323\\} &\tabincell{c} {5.688\\5.511\\5.530\\5.122\\5.325\\}&\tabincell{c}{6.262\\6.162\\6.234\\5.776\\6.167\\}   \\\hline
  \tabincell{c} {0.5*N(-6,3) \\+0.5*N(6,9)\\} & \tabincell{c} {1 \\ 0.5\\0\\-0.5\\ -1 \\}& \tabincell{c} {0.021\\0.018\\0.024\\-0.010\\-0.023   }&\tabincell{c} {0.253\\0.190\\0.175\\0.239\\0.283\\ }&\tabincell{c} {-0.206\\ -0.232\\-0.271 \\-0.263 \\-0.380 \\}&\tabincell{c} {0.350\\0.302\\ 0.344\\0.346\\0.481\\} &\tabincell{c} {4.930\\ 5.882\\5.379 \\5.022 \\ 5.757\\}&\tabincell{c}{5.862\\  6.584\\ 5.592\\ 5.590\\6.645 \\}   \\\hline

\end{tabular}
\label{DD}
\end{table}
\begin{table}
\newcommand{\tabincell}[2]{\begin{tabular}{@{}#1@{}}#2\end{tabular}}
 \renewcommand{\tabcolsep}{0.2pc}
\renewcommand{\arraystretch}{0.6}
  \centering
   \caption{Compared the proposed method with the CML when the error term is standard normal distribution. With the distribution of individual effect $\tau_i$ distribution is N(0,1),N(0,$\pi^2/3$), $N(0,6)$, $0.5*N(-4,6)+0.5*N(4,6)$ for sample n=500 based on 100 repetitions.}
	\begin{tabular}{ccccccccc}
		\hline
		  &  & \multicolumn{2}{c}{New method} &\multicolumn{2}{c}{ CML}\\
		\hline
 Distribution of the $\tau_i$& $\lambda $& BIAS$(\hat{\lambda}_{new})$ &RMSE$(\hat{\lambda}_{new})$& BIAS$(\hat{\lambda}_{CML})$ &RMSE$(\hat{\lambda}_{CML})$   \\ \hline
$N(0,1)$ &  \tabincell{c}{1\\0.5\\0\\-0.5\\-1\\} & \tabincell{c}{-0.011\\-0.006\\0.004\\-0.002\\-0.001\\}   & \tabincell{c}{0.119\\0.102\\0.083\\0.110\\0.118 \\} & \tabincell{c}{0.194\\0.077\\-0.002\\-0.088\\-0.205\\ } & \tabincell{c}{0.227\\0.131\\0.122\\0.134\\0.236\\ }    \\\hline
 $ N(0,\pi^2/3)$& \tabincell{c}{1\\0.5\\0\\-0.5\\-1\\} & \tabincell{c}{0.011\\0.006\\0.001\\0.002\\-0.029\\ } & \tabincell{c}{0.184\\0.140\\0.133\\0.139\\0.207\\ } & \tabincell{c}{-0.504\\-0.239\\-0.015\\0.238\\0.500\\} & \tabincell{c}{0.510\\0.248\\0.070\\0.247\\0.505\\}\\ \hline
  $N(0,6)$& \tabincell{c}{1\\0.5\\0\\-0.5\\-1\\} & \tabincell{c}{0.056\\0.010\\-0.007\\0.008\\-0.018\\ } &  \tabincell{c}{0.248\\0.199\\0.191\\0.199\\0.232\\ } & \tabincell{c}{-0.712\\-0.350\\0.008\\0.359\\0.716\\}  & \tabincell{c}{0.714\\0.354\\0.046\\0.363\\0.718 \\}\\ \hline
   $0.5*N(-4,6)+0.5*N(4,6)$& \tabincell{c}{1\\0.5\\0\\-0.5\\-1\\}& \tabincell{c}{0.019\\0.002\\-0.005\\-0.010\\-0.004\\ } & \tabincell{c} {0.269\\0.219\\0.202\\0.222\\0.249\\}& \tabincell{c}{-0.763\\-0.385\\0.008\\0.183\\0.761\\}  & \tabincell{c}{ 0.765\\0.385\\0.050\\0.386\\0.762 \\}\\ \hline
	\end{tabular}
\label{DE}
\end{table}

\par
When the error term is standard logistic distribution for models $(\ref{ME})$, Chamberlain (1980) proposed conditional maximum likelihood (CML) estimator for the parameter $\lambda$ and Bartolucci and Pigini (2017) developed an R package which is named as {\bf cquad} to implement the CML approach. Here, we compare the proposed method with the CML when the individual effects $\tau_i$ is distributed by N(0,1), $N(0,\pi^2/3)$, $N(0,6)$, $0.5*N(-4,6)+0.5*N(4,6)$, respectively, $\epsilon_i^A$ and $\epsilon_i^B $ are independently standard normal distribution and $D_i$ follows binomial distribution with success probability being $p=2/3$. Simulation results are listed in Table $\ref{DE}$ with sample n=500 and 100 repetitions. Simulation studies show that the biases of the CML are large, furthermore the biases are increasing with the increase of variance of individual effects.

\par
Finally, we compare the proposed estimator with the inverse probability weighted estimator developed by Robins {\it et al.} (1994). Here simulated data are generated by the following models:
 \begin{equation*}
Y^A_i=I({\beta^{'}\bx}_i^{A}+\tau_i+\lambda D_i+\epsilon^{A}_i>0), \mbox{~and}\;\;Y_i^B=I({\beta^{'}\bx}_i^{B}+\tau_i+\lambda (1-D_i)+\epsilon^{B}_i>0). \label{MZ}
\end{equation*}
$$D_i=I\{0.75\bx_i^{A}+0.25\bx_i^{B}+\varepsilon_i>0\}$$
where $\beta=1$, $ \bx_{i}^{A}$ being sampled from the standard normal distribution, $\bx_{i}^{B}$ being sampled from the uniform distribution $U(-1,1)$, and $\varepsilon_i$ following standard logistic distributions and the group effects being normally distributed with mean 0 and variance $\pi^2/3$.
BIASs and RMSEs are reported in Table $\ref{DF}$ with the sample size $n=1000$ and $100$ repetitions. It is shown that both the proposed estimators and Heckman's estimators perform better than IPW estimators. 

\begin{table}
\newcommand{\tabincell}[2]{\begin{tabular}{@{}#1@{}}#2\end{tabular}}
 \renewcommand{\tabcolsep}{0.1pc}
\renewcommand{\arraystretch}{0.6}
  \centering
   \caption{ Compared the proposed estimator $\hat{\lambda}_{new}$ with IPW estimator $\widehat{TE}_{IPW}$ and Heckman estimator $\hat{\lambda}_H$ for group effects being distributed as $N(0,\pi^2/3) $ and sample size being 1000 based on 100 repetitions.}
  \begin{tabular}{cccccccc}\hline
 $\lambda$ &BIAS$(\hat{\lambda}_{new})$ &RMSE$(\hat{\lambda}_{new})$  & BIAS$(\widehat{TE}_{IPW})$  &RMSE$(\widehat{TE}_{IPW})$ & BIAS$(\hat{\lambda}_H)$& RMSE$(\hat{\lambda}_H)$& RMSE$(\hat{\sigma}_H)$  \\ \hline
 \tabincell{c}{1 \\ 0.5 \\0\\ -0.5 \\ -1 \\} & \tabincell{c}{ 0.015\\-0.007\\-0.006\\0.003\\-0.022 \\ } & \tabincell{c} {0.159\\0.136\\0.119\\0.147\\0.163\\}& \tabincell{c}{-0.883\\-0.441\\0.004\\0.449\\0.894\\ } & \tabincell{c}{0.883 \\ 0.441 \\0.014 \\ 0.448 \\ 0.894\\}  & \tabincell{c}{0.030\\0.009\\0.001\\0.013\\-0.047 \\} &  \tabincell{c}{ 0.152\\0.110\\0.101\\0.120\\0.143\\ } & \tabincell{c}{0.410\\0.354\\0.354\\0.330\\0.385\\}   \\\hline
\end{tabular}
\label{DF}
\end{table}

\section{Applications}
In this section, we will demonstrate our methodologies with two  real datasets -  one is about lead levels in children's blood and the other is about acute leukaemia patients from a clinical trial.
\subsection{ Lead Levels in Children's Blood }
The data set shown in the Table $\ref{DG}$ are extracted from the observational study being studied in Morton {\it et al.} (1982), Pruzek \& Helmreich (2009). Briefly, children of parents who had worked in a lead-unrelated industry where lead was used in making batteries were matched by age, exposure to traffic, and neighborhood with children whose parents did not work in lead-related industries. We apply our proposed method to estimate the influence of environmental sources on children of lead absorption. here, we define that $Y$ = 1 if  lead level (ug/dl) exceeds 16, 0 otherwise. Our proposed method yields $\hat{\lambda}=0.9992$ with $\hat{\sigma}^2_{\lambda}  =0.1733$ while Heckman's method produces $\hat{\lambda}_H=1.2278$ with $\hat{\sigma}^2_H=0.1257$. According to our proposed method, the odd of treatment group and control group  is given by $P(Y^A=1\mid \tau)/P(Y^B=1 \mid \tau)=\int \Phi(\tau+0.9992)f(\tau)d\tau/\int \Phi(\tau)f(\tau)d\tau=1.653$ . Hence, children of employees in a battery manufacturing plant have  higher prevalence of high level of blood lead than those children whose parents are not employed in a lead-related industry.
\begin{table}
\newcommand{\tabincell}[2]{\begin{tabular}{@{}#1@{}}#2\end{tabular}}
 \renewcommand{\tabcolsep}{2.8pc}
\renewcommand{\arraystretch}{0.8}
  \centering
   \caption{Blood lead levels of case and control children and the difference between blood lead levels
of the two groups by matched pairs.}
  \begin{tabular}{cccc}\hline

  Matched & \multicolumn{3}{c}{Lead levels(ug/dl)} \\ \cline{2-4}
 pair no  &  case & control & Difference   \\  \hline
 \tabincell{c}{1\\2\\ 3\\4\\5\\6\\7\\8\\9\\10\\11\\12\\13\\14\\15\\16\\17\\18\\19\\20\\21\\22\\23\\24\\25\\26\\27\\28\\29\\30\\31\\32\\33\\} &
 \tabincell{c}{38 \\23 \\41 \\18 \\37 \\36\\ 23 \\62 \\31 \\34 \\24\\ 14 \\21 \\17 \\ 16 \\ 20 \\ 15 \\ 10\\45\\39\\22\\35\\49\\48\\44\\35\\43 \\39\\34 \\13\\73\\25\\27\\} &
 \tabincell{c}{16\\18\\18\\24\\19\\11\\10\\15\\16\\18\\18\\13\\19\\10 \\16 \\16 \\24 \\13 \\ 9 \\14 \\21 \\19 \\ 7 \\18 \\19 \\12 \\11\\ 22\\25 \\16\\ 13\\ 11\\ 13\\} &
  \tabincell{c}{22\\5 \\23\\ -6 \\18\\ 25\\ 13\\ 47\\ 15\\ 16 \\ 6 \\ 1 \\ 2 \\ 7 \\ 0 \\ 4 \\-9 \\-3 \\36\\ 25 \\1\\16\\42\\30\\25\\23\\32\\ 17\\9\\ -3\\ 60\\ 14\\ 14\\}\\ \hline
 mean  & 31.84848 &15.87879 & \\ \hline
\end{tabular}
\label{DG}
\end{table}

 \subsection{ Acute leukaemia patients of clinical trial}

The data reported in Table $\ref{DH}$ are coming from a clinical trial about acute leukaemia patients, see Freireich {\it et al.} (1963), Gehan (1965). Briefly, treatment group which uses 6-mercaptopurine (6-MP) to therapy is compared to control group which employs placebo to treatment in the maintenance of remissions in acute leukemia. One year after the start of the study, times of remission in weeks are  recorded. We define $Y$ = 1 if the time of remission is more than 12 weeks, 0  otherwise. Our proposed method yields $\hat{\lambda}=0.617$ with $\hat{\sigma}^2_{\lambda}  =0.1377$ while Heckman's method produces $\hat{\lambda}_H=0.180$ with $\hat{\sigma}^2_H=0.1390$. Obviously, simple Wald tests will suggest significant treatment effect from our proposed method and insignificant treatment effect from Heckman's method. According to our method, it can be concluded that the odd of treatment group and control group is $ P(Y^A=1 \mid \tau)/P(Y^B=1 \mid \tau)=\int \Phi(\tau+0.617)f(\tau)d\tau/\int \Phi(\tau)f(\tau)d\tau= 1.337$. Consequently, there is  strong evidence that patients receiving 6-MP have longer remissions than those receiving placebo.
\begin{table}
\newcommand{\tabincell}[2]{\begin{tabular}{@{}#1@{}}#2\end{tabular}}
 \renewcommand{\tabcolsep}{2.0pc}
\renewcommand{\arraystretch}{0.8}
  \centering
   \caption{Length of remission of 42 acute leukaemia patients with 21 patients being treated with the drug 6-mercaptopurine and 21 being assigned to control.}
  \begin{tabular}{cccc}\hline
 Pair &Remission time(weeks) & Censoring  & Group \\  \hline
 \tabincell{c}{1\\1 \\2\\2\\3\\3\\4\\4\\5\\5\\6\\6\\7\\7\\8\\8\\9\\9\\10\\10\\11\\11\\12 \\12 \\13\\13\\14\\14\\15\\15\\16 \\16\\17\\ 17\\ 18 \\18 \\19 \\19 \\20 \\20 \\21\\21\\} &
 \tabincell{c}{1\\10\\ 22\\7\\3\\32\\12\\ 23\\8\\22\\17\\ 6\\2\\16\\11\\34\\8\\32\\12\\25\\2\\11\\5 \\20 \\4 \\19\\15\\6\\8\\17\\23 \\35 \\5\\ 6 \\11 \\13 \\4\\9\\1\\6\\ 8\\10\\} &
 \tabincell{c}{ 1\\ 1\\ 1\\ 1\\1\\ 0\\ 1\\ 1\\1\\1\\1\\1\\ 1\\ 1\\ 1\\ 0\\ 1\\ 0\\ 1\\ 0\\ 1\\ 0\\ 1\\ 0\\ 1\\ 0 \\1\\ 1\\ 1\\ 0 \\1 \\0 \\1\\ 1\\ 1\\ 1\\ 1\\ 0\\ 1 \\ 0\\1\\ 0\\} &
  \tabincell{c}{control\\6-MP\\control\\6-MP\\control\\6-MP\\control\\6-MP\\control\\6-MP\\control\\6-MP\\control\\6-MP\\control\\6-MP\\control\\6-MP\\control\\6-MP\\
  control\\6-MP\\control\\6-MP\\control\\6-MP\\control\\6-MP\\control\\6-MP\\control\\6-MP\\control\\6-MP\\control\\6-MP\\control\\6-MP\\control\\6-MP\\control\\6-MP\\  }\\ \hline
\end{tabular}
\label{DH}
\end{table}
\par

\section{Conclusion}
 \par
In this paper, we consider a matched pairs data design which allows for unobserved individual heterogeneity (groups effects) with treatment effects on the dependent variable. A new method for estimating treatment effects based on the probit model is proposed. Here, group effects with large variances are allowed. We demonstrate that our proposed parameter estimators are consistent and asymptotically normally distributed. In general, our proposed estimators are superior to existing Heckman's estimators and  inverse probability weighted (IPW) estimators.

\section*{Appendix. Proofs}

We assume that  $(\bx^{A}_{i},\bx^{B}_{i},D_i)(i=1,\cdots,n)$ are mutually independent with a common distribution function $F(\bx^{A},\bx^{B},D)$. Let $\sigma_{\tau}=an^{\alpha}$ with $a>0$ and $0<\alpha<1$.

\par
{\bf Lemma A1}.  Under Assumption (\ref{AS1}) with $h(x)$ having continuous derivatives,  we have
\begin{eqnarray*}
\frac{1}{n^{1-\alpha}}\sum\limits_{i=1}^{n}I(Y_{i}^{A}+Y_{i}^{B}=1) \xrightarrow{\ p \ } C_{1},
\end{eqnarray*}
where $\displaystyle C_{1}=\frac{ h(0)}{a}E [K(\btheta;\bx^A,\bx^B,D)]$.

\par
{\bf Proof:}
\par
Letting $W_{i}=I(Y_{i}^{A}+Y_{i}^{B}=1)$ yields
\begin{eqnarray*}
E(W_{i}|\bx^A_i,\bx^B_i,D_i)&=&\bigg[\int \Phi(\bbeta^{'} \bx^A_i+\lambda D_i + u)\Phi(-\bbeta^{'}\bx^B_i - \lambda(1- D_i)-u)\frac{1}{\sigma_{u}}h(\frac{ u-\mu_{\tau}}{\sigma_{\tau}})du   \\
&&+ \int \Phi (-\bbeta^{'} \bx^A_i-\lambda D_i -u)\Phi(\bbeta^{'}\bx^B_i +\lambda(1- D_i)+u) \frac{1}{\sigma_{\tau}}h(\frac{u-\mu_{\tau}}{\sigma_{\tau}})du\bigg]  \\
&=&\frac{1}{an^{\alpha}}\times\bigg[\int \Phi (\bbeta^{'} \bx^A_i+\lambda D_i+u)\Phi(-\bbeta^{'}\bx^B_i -\lambda(1- D_i)-u)
 h(\frac{ u-\mu_{\tau}}{\sigma_{\tau}})du  \\
&&+ \int \Phi(-\bbeta^{'} \bx^A_i-\lambda D_i-u)\Phi(\bbeta^{'}\bx^B_i +\lambda(1- D_i) +u)
 h(\frac{ u-\mu_{\tau}}{\sigma_{\tau}})du \bigg]   \\
&=&\frac{h(0)}{an^{\alpha}}\times\bigg[\int \Phi(\bbeta^{'} \bx^A_i+\lambda D_i +u )\Phi (-\bbeta^{'}\bx^B_i -\lambda(1- D_i) -u)du  \\
&&+ \int \Phi(-\bbeta^{'} \bx^A_i-\lambda D_i-u )\Phi(\bbeta^{'}\bx^B_i +\lambda(1- D_i) +u)du +O(\sigma_{\tau}^{-1}) \bigg]  \\
&=&\frac{h(0)}{an^{\alpha}}[K(\btheta;\bx^A_i,\bx^B_i,D_i)+O(n^{-\alpha}) ]
\end{eqnarray*}
and
$$
E(W_i)=\frac{C_1}{n^\alpha}+O(n^{2\alpha}),\;\;\frac{1}{n^{1-\alpha}}\sum_{i=1}^{n}E(W_{i})=C_1+O(n^{-\alpha})=C_1+o(1).
$$
Therefore,
$$
E(W_{i})^2 =E W_{i} = \frac{C_{1}}{n^{\alpha}}+O(n^{-2\alpha}),\;\;  Var(W_i)=E(W_i)^2-(E(W_i))^2=O(n^{-\alpha})
$$
and
$$
 Var(\frac{1}{n^{1-\alpha}}\sum_{i=1}^{n}W_{i})^2=\frac{1}{n^{2-2\alpha}}nVar(W_{i})^2=\frac{1}
 {n^{2-2\alpha}}n(\frac{C_{1}}{n^{\alpha}}+O(n^{-2\alpha}))=O(n^{1-\alpha}).
$$
By Chebyshev's inequality, we  have
\begin{eqnarray*}
P\bigg(\bigg|\frac{1}{n^{1-\alpha}}\sum\limits_{i=1}^{n}W_{i}-E(\frac{1}{n^{1-\alpha}}\sum\limits_{i=1}^{n}W_{i})\bigg|\geq \varepsilon\bigg)
&=&P\bigg(\bigg|\frac{1}{n^{1-\alpha}}\sum\limits_{i=1}^{n}W_{i} -(C_{1}+o(1))\bigg|\geq \varepsilon\bigg) \\
& \leq & \frac{Var\left(\frac{1}{n^{1-\alpha}}\sum\limits_{i=1}^{n}W_{i}\right)^2}{\varepsilon^{2}} \longrightarrow 0,
\end{eqnarray*}
which implies Lemma A1.
\par
{\bf Lemma A2}. Under Assumption (\ref{AS1}), we have
\begin{equation*}
\frac{1}{n^{1-\alpha}}\sum\limits_{i=1}^{n}I(Y_{i}^{A}+Y_{i}^{B}=1)\left\{{Y_{i}^{A}\log p({\btheta}^{*}; \bx^{A}_i,\bx^{B}_i,D_i)}+{Y_{i}^{B}\log(1-p({\btheta}^{*}; \bx^{A}_i,\bx^{B}_i,D_i))}\right\} \xrightarrow{p}C_{2},
\end{equation*}
where
\begin{eqnarray*}
C_{2}({\btheta}^{*})&=&\frac{h(0)}{a}E \bigg\{K(\btheta; \bx^A,\bx^B,D)\bigg[p({\btheta};\bx^{A},\bx^{B},D)\log p({\btheta}^{*};\bx^{A},\bx^{B},D) \\
&&+(1-p({\btheta};\bx^A,\bx^B,D))\log(1-p({\btheta}^{*};\bx^A,\bx^B,D))\bigg]\bigg\}.
\end{eqnarray*}
\par
{\bf Proof:} Let
\begin{equation*}
U_{i}=I(Y_{i}^{A}+Y_{i}^{B}=1)\left\{{Y_{i}^{A}\log p({\btheta}^{*};\bx^A_{i},\bx^B_{i},D_i)}+ {Y_{i}^{B}\log(1-p({\btheta}^{*};\bx^A_{i},\bx^B_{i},D_i))}\right\}.
\end{equation*}
We have
\begin{eqnarray*}
E(U_{i})&=&E\left[E(U_{i}\mid x^A_{i},x_{i}^B,D_i)\right] \\
&=&\dfrac{h(0)}{a  n^{\alpha}}\Big\{E\Big[\int \Phi(\bbeta^{'} \bx^A+\lambda D+u)\Phi(-\bbeta^{'} \bx^B -\lambda (1-D)-\tau)du\log p({\btheta}^{*};\bx^A,\bx^B,D)\\
&&+{\int \Phi(-\bbeta^{'} \bx^A-\lambda(1-D)-u)\Phi(\bbeta^{'}\bx^B+\lambda D +u)du}\log(1-p({\btheta}^{*};\bx^A,\bx^B,D))\Big]\\
&&\;\;\;+o(1)\Big\}\\
&=&\frac{1}{n^{\alpha}}\{C_{2}({\btheta}^{*})+o(1)\}
\end{eqnarray*}
and
\begin{equation*}
\frac{1}{n^{1-\alpha}}E\left(\sum\limits_{i=1}^{n}U_{i}\right)=\frac{1}{n^{1-\alpha}}\sum\limits_{i=1}^{n}
\frac{1}{n^{\alpha}}\{C_{2}({\btheta}^{*})+o(1)\}=C_{2}({\btheta}^{*})+o(1).
\end{equation*}
\par
Similarly, one can obtain
\begin{eqnarray*}
E(U^{2}_{i})&=&E\left[E(U^{2}_{i}|\bx^A_{i},\bx^B_{i},D_i)\right]\\
&=&\dfrac{h(0)}{a  n^{\alpha}}\Big\{E\Big[{\log^{2}p({\btheta};\bx^A,\bx^B,D)\int \Phi({\bbeta}^{'}\bx^A +\lambda D +u) \Phi(-\bbeta^{'} \bx^B -\lambda(1-D)-u)du}\\
&&+{\log^{2}(1-p({\btheta};\bx^A,\bx^B,D))\int \Phi(-\bbeta^{'}\bx^A-\lambda D-u)\Phi(\bbeta^{'}\bx^B+\lambda(1-D)+u)du} \Big]+o(1)\Big\}\\
&=&O(n^{-\alpha}),
\end{eqnarray*}
\begin{equation*}
Var(U_{i})=E(U^{2}_{i})-[E(U_{i})]^{2}=O(n^{-\alpha}), \mbox{~and}
\end{equation*}

\begin{equation*}
Var\left(\frac{1}{n^{1-\alpha}}\sum\limits_{i=1}^{n}U_{i}\right) =\frac{1}{n^{2(1-\alpha)}}\sum\limits_{i=1}^{n}Var(U_{i})=O(n^{1-\alpha}).
\end{equation*}

\par
For $\varepsilon>0$, as $n\rightarrow \infty$  we have
\begin{eqnarray*}
P\bigg(\bigg|\frac{1}{n^{1-\alpha}}\sum\limits_{i=1}^{n}U_{i}-E(\frac{1}{n^{1-\alpha}}\sum\limits_{i=1}^{n}U_{i})\bigg|\geq \varepsilon\bigg)
&=&P\bigg(\bigg|\frac{1}{n^{1-\alpha}}\sum\limits_{i=1}^{n}U_{i} -(C_{2}+o(1))\bigg|\geq \varepsilon\bigg)\\
&\leq&\frac{Var\left(\frac{1}{n^{1-\alpha}}\sum\limits_{i=1}^{n}U_{i}\right)}{\varepsilon^{2}} \longrightarrow 0
\end{eqnarray*}
which implies that Lemma A2 holds.
\par
{\bf Proof of Theorem 2:} Let
$$
L({\btheta}^{*})=\prod\limits_{i=1}^{n}p({\btheta}^{*};\bx^A_{i},\bx^B_{i},D_i)^{Y_{i}^{A}(1-Y_{i}^{B})}(1-p({\btheta}^{*};\bx^A_{i},\bx^B_{i},
D_i))^{Y_{i}^{B}{(1-Y_{i}^{A})}}
$$
 and its logarithm be
$$
l({\btheta}^{*})=\sum\limits_{i=1}^{n}I(Y_{i}^{A}+Y_{i}^{B}=1)\left\{{Y_{i}^{A}\log p({\btheta}^{*};\bx^A_{i},\bx^B_{i},D_i)}+{Y_{i}^{B}\log(1-p({\btheta}^{*};\bx^A_{i},\bx^B_{i},D_i))}\right\}.
$$
By Lemmas A1 and A2, we have
$$
\frac{1}{n^{1-\alpha}}l({\btheta}^{*})=\frac{1}{n^{1-\alpha}}\sum\limits_{i=1}^{n}U_{i} \xrightarrow{\ p \ }C_{2}({\btheta}^{*})
$$
and
$$
\frac{1}{n^{1-\alpha}}\sum\limits_{i=1}^{n} I(Y_{i}^{A}+Y_{i}^{B}=1) \xrightarrow{\ p \ }C_{1}.
$$
Thus,
$$
\frac{1}{\sum\limits_{i=1}^{n}I(Y_{i}^{A}+Y_{i}^{B}=1)}l({\btheta}^{*})\xrightarrow{\ p \ }\frac{C_2({\btheta}^{*})}{C_1}.
$$

\par
Since
$$
p({\btheta};\bx^A,\bx^B,D)\log \frac{p({\btheta};\bx^A,\bx^B,D)}{p({\btheta}^{*};\bx^A,\bx^B,D)}
+(1-p({\btheta};\bx^A,\bx^B,D)) \log\frac{(1-p({\btheta};\bx^A,\bx^B,D))}{(1-p({\btheta}^{*}; \bx^A,\bx^B,D))}\geq 0,
$$
we have
$$
P\left\{\frac{1}{\sum\limits_{i=1}^{n}I(Y_{i}^{A}+Y_{i}^{B}=1)}l({\btheta}^{*})\leq
\frac{1}{\sum\limits_{i=1}^{n}I(Y_{i}^{A}+Y_{i}^{B}=1)}l({\btheta})\right\}\xrightarrow{\ p \ }1,
$$
 which implies that $\hat{\btheta}$  is a consistent estimator of ${\btheta}$.
\par
{\bf Lemma A3}. Under Assumption (\ref{AS1}) with $h(x)$ having continuous derivatives and $\sigma_{\tau}=a\sqrt{n} (a>0)$,
\begin{equation*}
\sum\limits_{i=1}^{n}\frac{ I(Y_{i}^{A}+Y_{i}^{B}=1)}{\sqrt{\sum\limits_{j=1}^{n}I(Y_{j}^{A}+Y_{j}^{B}=1)}} \left\{\frac{Y_{i}^{A}-p({\btheta};\bx^A_i,\bx^B_i,D_i)}
{p({\btheta};\bx^A_i,\bx^B_i,D_i)(1-p({\btheta};\bx^A_i,\bx^B_i,D_i))} \frac{\partial p({\btheta};\bx^A_i,\bx^B_i,D_i)}{\partial {\btheta}}\right\}
\end{equation*}
has the asymptotic distribution $N(0,\Sigma/c)$, where
$$
\Sigma=\displaystyle E\left\{\frac{K(\btheta;\bx^A,\bx^B,D) \Big(\frac{\partial  p({\btheta};\bx^A,\bx^B,D)} {\partial {\btheta}} \Big)\Big(\frac{\partial  p({\btheta};\bx^A,\bx^B,D)} {\partial {\btheta}} \Big)^{'}}{p({\btheta};\bx^A,\bx^B,D)(1-p({\btheta};\bx^A,\bx^B,D))}\right\}, \;\;c=E[K(\btheta;\bx^A,\bx^B,D)].
$$
\par
{\bf Proof: } Let
\begin{eqnarray*}
Q_{i}&=&I(Y_{i}^{A}+Y_{i}^{B}=1)\Big\{\frac{Y_{i}^{A}-p({\btheta};\bx^A_i,\bx^B_i,D)}
{p({\btheta};\bx^A_i,\bx^B_i,D_i)(1-p({\btheta};\bx^A_i,\bx^B_i,D_i))} \frac{\partial p({\btheta};\bx^A_i,\bx^B_i,D_i)} {\partial {\btheta}}\Big\} \\\\
&=&  \Big\{ \frac{I(Y_{i}^{A}+Y_{i}^{B}=1) Y_i^A}{p({\btheta};\bx^A_i,\bx^B_i,D_i)} -\frac{I(Y_{i}^{A}+Y_{i}^{B}=1) Y_i^B}{1-p({\btheta};\bx^A_i,\bx^B_i,D_i)}\Big\} \frac{\partial p({\btheta};\bx^A_i,\bx^B_i,D_i)} {\partial {\btheta}}.
\end{eqnarray*}
Since the first derivative of  $h(x)$ is continuous, we have
\begin{eqnarray*}
&&E\left(\frac{I(Y_{i}^{A}+Y_{i}^{B}=1) Y_i^A}{p({\btheta};\bx^A_i,\bx^B_i,D_i)} \mid \bx^A_i,\bx^B_i,D_i\right)\\
&&=\frac{\int \Phi(\bbeta^{'} \bx^A_i+\lambda D_i +u)\Phi(-\bbeta^{'} \bx^B_i-\lambda(1-D_i)-u)h(\frac{u-\mu_{\tau}}{\sigma_{\tau}})du}{\sigma_\tau\times p({\btheta};\bx^A_i,\bx^B_i,D_i)}\\
&&=\frac{\int \Phi(\bbeta^{'} \bx^A_i+\lambda D_i +u)\Phi(-\bbeta^{'} \bx^B_i-\lambda(1-D_i)-u)[h(0)+h^{'}(0)\frac{u-\mu_{\tau}}{\sigma_{\tau}} +o(\sigma_{\tau}^{-1})]du}{\sigma_\tau \times p({\btheta};\bx^A_i,\bx^B_i,D_i)}\\
&=&\frac{h(0)\int \Phi(\bbeta^{'} \bx^A_i+\lambda D_i +u)\Phi(-\bbeta^{'} \bx^B_i-\lambda(1-D_i)-u)du}{\sigma_\tau \times p({\btheta};\bx^A_i,\bx^B_i,D_i)}+O(\sigma^{-2}_\tau)+o(\sigma^{-2}_\tau)
\end{eqnarray*}
and
\begin{eqnarray*}
&&E\left(\frac{I(Y_{i}^{A}+Y_{i}^{B}=1) Y_i^B}{1-p({\btheta};\bx^A_i,\bx^B_i,D_i)}\mid \bx^A_i,\bx^B_i,D_i\right)\\
&&=\frac{\int \Phi(-\bbeta^{'} \bx^A_i-\lambda D_i -u)\Phi(\bbeta^{'} \bx^B_i+\lambda(1-D_i)+u)h(\frac{u-\mu_{\tau}}{\sigma_{\tau}})du}{\sigma_\tau\times [1-p({\btheta};\bx^A_i,\bx^B_i,D_i)]}\\
&&=\frac{\int \Phi(-\bbeta^{'} \bx^A_i-\lambda D_i -u)\Phi(\bbeta^{'} \bx^B_i+\lambda(1-D_i)+u)[h(0)+h^{'}(0)\frac{u-\mu_{\tau}}{\sigma_{\tau}} +o(\sigma_{\tau}^{-1})]du}{\sigma_\tau \times [1-p({\btheta};\bx^A_i,\bx^B_i,D_i)}]\\
&=&\frac{h(0)\int \Phi(-\bbeta^{'} \bx^A_i-\lambda D_i -u)\Phi(\bbeta^{'} \bx^B_i+\lambda(1-D_i)+u)du}{\sigma_\tau \times [1-p({\btheta};\bx^A_i,\bx^B_i,D_i)]}+O(\sigma^{-2}_\tau)+o(\sigma^{-2}_\tau).
\end{eqnarray*}
As
$$
\frac{\int \Phi(\bbeta^{'} \bx^A_i+\lambda D_i +u)\Phi(-\bbeta^{'} \bx^B_i-\lambda(1-D_i)-u)du}{\int \Phi(-\bbeta^{'} \bx^A_i-\lambda D_i -u)\Phi(\bbeta^{'} \bx^B_i+\lambda(1-D_i)+u)du }=\frac{p({\btheta};\bx^A_i,\bx^B_i,D_i)}{ 1-p({\btheta};\bx^A_i,\bx^B_i,D_i)},
$$
we have
$$
E(Q_i)=E(E(Q_{i} \mid \bx^A_i,\bx^B_i,D_i))
=O(\sigma^{-2}_\tau)+o(\sigma^{-2}_{\tau}) =O(\sigma^{-2}_\tau)=O(n^{-1}).
$$
Similarly,
\begin{eqnarray*}
E[Q_{i}Q_{i}^{'}]&=&\frac{h(0)}{a\sqrt{n}}\times E\left\{\frac{K(\btheta;\bx^A,\bx^B,D)\Big(\frac{\partial p({\btheta};\bx^A,\bx^B,D)} {\partial {\btheta}}\Big)\Big(\frac{\partial p({\btheta};\bx^A,\bx^B,D)} {\partial {\btheta}}\Big)^{'}}{[1-p({\btheta};\bx^A,\bx^B,D)]p({\btheta};\bx^A,\bx^B,D)}\right\}\\
&&+o(n^{-1/2})\\
&=&\frac{h(0)}{a\sqrt{n}}\times \Sigma+o(n^{-1/2}).
\end{eqnarray*}
\par
Let ${\bf{t}}=(t_{1},\cdots,t_{k+1})^{'}$ be a k+1 dimensional row vector. We have
\begin{eqnarray*}
E[\exp\{i{\bf{t}}^{'}Q_{j}/n^{1/4}\}]&=&1+\frac{1}{n^{1/4}}E(i{\bf{t}}^{'}Q_{j})-\frac{1}{2\sqrt{n}}E(-{\bf{t}}^{'}Q_{j}Q^{'}_{j}{\bf{t}})
+E\Big[o\Big(\frac{{\bf{t}}^{'}Q_{j}Q^{'}_{j}{\bf{t}}}{n^{1/2}}\Big)\Big]\\
&=&1-\frac{1}{2\sqrt{n}}E({\bf{t}}^{'}Q_{j}Q^{'}_{j}{\bf{t}})+o(n^{-1})\\
&=&1-\frac{h(0){\bf{t}}^{'}\Sigma{\bf{t}}}{2an}+o(n^{-1}).\\
\end{eqnarray*}
The characteristic function of $\sum\limits_{j=1}^{n}Q_{j}/n^{1/4}$ is given by
\begin{eqnarray*}
\varphi_{n}({\bf{t}})&=&E\Big[\exp\{i\sum\limits_{j=1}^{n}{\bf{t}}^{'}Q_{j}/n^{1/4}\}\Big]
=\Big[E(\exp\{i{\bf{t}}^{'}Q_{j}/n^{1/4}\})\Big]^{n}\\
&=&\Big[1-\frac{h(0){\bf{t}}^{'}\Sigma{\bf{t}}}{2an}+o(n^{-1})\Big]^{n}
\longrightarrow \exp\Big[-\frac{h(0){\bf{t}}^{'}\Sigma{\bf{t}}}{2a}\Big].
\end{eqnarray*}
Hence,
\begin{equation*}
\frac{1}{n^{1/4}}\sum\limits_{j=1}^{n}Q_{j}\xrightarrow{\ d \ }N(0,\frac{h(0)}{a}\Sigma).
\end{equation*}
By Lemma A1, we have
$$
\frac{1}{\sqrt{n}}\sum\limits_{i=1}^{n} I(Y_{i}^{A}+Y_{i}^{B}=1)\xrightarrow{\ p \ }
\frac{ h(0)}{a}E[ K(\btheta;\bx^A,\bx^B,D)].
$$
As a result,
\begin{eqnarray*}
\sum\limits_{i=1}^{n}\frac{I{(Y_{i}^{A}+Y_{i}^{B}=1)}}{\sqrt{\sum\limits_{j=1}^{n} I{(Y_{j}^{A}+Y_{j}^{B}=1)}}}
 \big[\frac{Y_{i}^{A}-p({\btheta};\bx^A_i,\bx^B_i,D_i)}
{p({\btheta};\bx^A_i,\bx^B_i,D_i)(1-p({\btheta};\bx^A_i,\bx^B_i,D_i))}
\frac{\partial p({\btheta};\bx^A_i,\bx^B_i,D_i)} {\partial {\btheta}}\big]\xrightarrow{\ d  \ }N(0,\Sigma/c)
\end{eqnarray*}
which implies the Lemma A3 holds.
\par
{\bf Proof of Theorem 3:} Let
$$
l(\btheta)=\sum\limits_{i=1}^{n}Y_i^{A}(1-Y_i^{B})\log p(\btheta;\bx^A_i,\bx^B_i,D_i)+(1-Y_i^{A})Y_i^{B}\log(1- p(\btheta;\bx^A_i,\bx^B_i,D_i))
$$
$\hat\btheta$ is its maximum point which satisfies
$$
\frac{\partial l(\hat{\btheta})}{\partial{\btheta}}=\sum\limits_{i=1}^{n}I(Y_{i}^{A}+Y_{i}^{B}=1)\left\{\frac{Y_{i}^{A}-p(\hat{\btheta};\bx^A_i,\bx^B_i,D_i)}
{p(\hat{\btheta};\bx^A_i,\bx^B_i,D_i)(1-p(\hat{\btheta};\bx^A_i,\bx^B_i,D_i))} \frac{\partial p(\hat{\btheta};\bx^A_i,\bx^B_i,D_i)}
{\partial {\btheta}}\right\}=0.
$$
Expanding the above function around the true value, i.e., ${\btheta}$, yields
\begin{eqnarray*}
\dfrac{\partial l(\hat{\btheta})}{\partial{\btheta}}
&=&\sum\limits_{i=1}^{n}I(Y_{i}^{A}+Y_{i}^{B}=1)\left\{\frac{Y_{i}^{A}-p({\btheta};\bx^A_i,\bx^B_i,D_i)}
{p({\btheta};\bx^A_i,\bx^B_i,D_i)(1-p({\btheta};\bx^A_i,\bx^B_i,D_i))}\frac{\partial p({\btheta};\bx^A_i,\bx^B_i,D_i)}
{\partial {\btheta}}\right\}\\
&&+\sum\limits_{i=1}^{n}I(Y_{i}^{A}+Y_{i}^{B}=1)\frac{\partial\left[\frac{Y_{i}^{A}-p({\btheta};\bx^A_i,\bx^B_i,D_i)}
{p({\btheta};\bx^A_i,\bx^B_i,D_i)(1-p({\btheta};\bx^A_i,\bx^B_i,D_i))}\frac{\partial p({\btheta};\bx^A_i,\bx^B_i,D_i)}
{\partial {\btheta}}\right]}{\partial {\btheta}}_{|{\btheta}={\btheta}}
(\hat{\btheta}-{\btheta}) \\
&& +[\sum\limits_{i=1}^{n}I(Y_{i}^{A}+Y_{i}^{B}=1)]o(\parallel \hat{\btheta}-{\btheta} \parallel).
\end{eqnarray*}

\par
By Lemma A3, it can be proved that
$$
\frac{1}{\sum\limits_{j=1}^{n}I(Y_{j}^{A}+Y_{j}^{B}=1)}\sum\limits_{i=1}^{n}I(Y_{i}^{A}+Y_{i}^{B}=1)
\frac{\partial\left[\frac{Y_{i}^{A}-p({\btheta};\bx^A_i,\bx^B_i,D_i)}
{p({\btheta};\bx^A_i,\bx^B_i,D_i)(1-p({\btheta};\bx^A_i,\bx^B_i,D_i))}\cdot\frac{\partial p({\btheta};\bx^A_i,\bx^B_i,D_i)}{\partial {\btheta}}\right]}{\partial{\btheta}}
$$
converges in probability to $-\Sigma/c$ and
\begin{equation*}
\sum\limits_{i=1}^{n}\frac{I(Y_{i}^{A}+Y_{i}^{B}=1) }{\sqrt{\sum\limits_{j=1}^{n}I(Y_{j}^{A}+Y_{j}^{B}=1)}}
\Big\{\frac{Y_{i}^{A}-p({\btheta};\bx^A_i,\bx^B_i,D_i)} {p({\btheta};\bx^A_i,\bx^B_i,D_i)(1-p({\btheta};\bx^A_i,\bx^B_i,D_i))}
\frac{\partial p({\btheta};\bx^A_i,\bx^B_i,D_i)} {\partial {\btheta}}\Big\}\xrightarrow{d} N(0, \Sigma/c)
\end{equation*}
Thus, one can  obtain
\begin{eqnarray*}
\sqrt{\sum\limits_{i=1}^{n}I(Y_{i}^{A}+Y_{i}^{B}=1)}(\hat{{\btheta}}-{\btheta})\xrightarrow{\ d \ } N(0, c\Sigma^{-1})
\end{eqnarray*}
and Theorem 3 is proved.

\section*{References}
\begin{description}
\item Abadie, A., Gardeazabal, J., 2003. The economic costs of conflict: a case study of the basque country. {\sl The American Economic Review.} {\bf 93}(1), 113--132.
\item Abadie, A., Imbens, G.W., 2006. Large sample properties of matching estimators for average treatment effects. { \sl Econometrica. } {\bf 74}, 235--267.
\item Abadie, A., Imbens, G.W., 2011. Bias-corrected matching estimators for average treatment effects. { \sl Journal of Business and Economic Statistics. } {\bf 29}, 1--11.
\item  Arellano, M., 2003. Discrete choices with panel data. {\sl Investigaciones Economicas.} {\bf27}(3), 423--458.
\item  Arellano, M., Bonhomme, S., 2009. Robust priors in nonlinear panel data models. {\sl The Econometric Society.} {\bf77}(2), 489--536.
\item Arulampalam, W., Stewart, M.B., 2009. Simplified implementation of the Heckman estimator of the dynamic probit model anda comparison with alternative estimators. { \sl Oxford Bulletin of Economics and Statistics.} {\bf 71}, 659--681.
\item Ashenfelter, O., Card, D., 1985. Using the longitudinal structure of earnings to estimate the effect of training programs. {\sl The Review of Economics and Statistics. } {\bf 67}(4), 648--660.
\item Bartolucci, F., Bellio, R., Salvan, A., Sartori, N., 2016. Modified profile likelihood for fixed-effects panel data models. {\sl Econometric Reviews.} {\bf35}(7), 1271--1289.
\item Bartolucci, F., Pigini, C., 2017. {\bf cquad}: An R and Stata package for conditional maximum likelihood estimation of dynamic binary panel data models. {\sl Journal of Statistical Software.} {\bf 78}, 1--26.
\item Bertschek, I., Lechner, M., 1998. Convenient estimators for the panel probit model. {\sl Journal of Econometrics.} {\bf 87}(2), 329--371.
 \item Butler, I., Moffitt, R., 1982. A computationally efficient quadrature procedure for the one-factor multinomial probit model. {\sl Econometrica.} {\bf 50}(3), 761--764.
\item Card, D., Krueger, A.B., 1994. Minimum wages and employment: a case study of the fast-food industry in New Jersey and Pennsylvania. {\sl American Economic Association.} {\bf 84}(4), 772--793.
\item Carro, J.M., 2007. Estimating dynamic panel data discrete choice models with fixed effects. {\sl Journal of Econometrics.} {\bf 140}(2), 503--528.
\item  Chamberlain, G., 1980. Analysis of covariance with qualitative data. {\sl The Review of Economic Studies.} {\bf 47}(1), 225--238.
\item Dey, K.D., Chen, M,H., Chang, H., 1997. Bayesian approach for nonlinear random effects models. {\sl  Biometrics.} {\bf 53}, 1239--1252.
\item D\'{i}az, J., Rau, T., Rivera, J., 2015. A matching estimator based on a bi-level optimization problem. {\sl Review of Economics and Statistics.} {\bf 97}(4), 803--812.
\item Freireich, E.J., Gehan, E.A., {\it et al.}, 1963. The effect of 6-mercaptopurine on the duration of steroid-induced remissions in acute leukemia. {\sl Blood.} {\bf 21}(6), 699--716.
 \item Gao, W., Bergsma, W., Yao, Q., 2017. Estimation for dynamic and static panel probit models with large individual effects. {\sl Jouranl of Time Series Analysis.} {\bf 38}, 266-284.
 \item Gehan, E., 1965. A generalized Wilcoxon test for comparing arbitrarily single-censored samples. {\sl Biometrika.} {\bf 52}, 203--233.
 \item Guilkey, D.K., Murphy, J.L., 1993. Estimation and testing in the random effects probit model. {\sl Journal of Econometrics.} {\bf 59}(3), 301--317.
\item  Heckman, J., 1978. Dummy endogenous variables in a simultaneous equation system. {\sl Econometrica.} {\bf 46}(4), 931--958.
\item  Heckman, J., Smith, J., Clements, N., 1997. Making the most out of programme evaluations and social experiments: accounting for heterogeneity in programme impacts. {\sl The Review of Economic Studies.} {\bf 64}(4), 487-535.
\item Hernan, M.A., Robins, J.M., 2006. Estimating Causal Effects From Epidemiological Data. {\sl Journal of Epidemiology and Community Health.} {\bf60}, 578¨C-596.
\item Hirano, K., Imbens, G.M., Ridder, G., 2003. Efficient estimation of average treatment effects using the estimated propensity score. {\sl  Econometrica.} {\bf 71}, 1161--1189.
\item Horowitz, J.L., 1992. A smoothed maximum score estimator for binary response model. {\sl Econometrica.} {\bf 60}, 505--531.
 \item Imbens, G.M., Angrist, D.J., 1994. Identification and estimation of local average treatment effects. {\sl Econometrica.} {\bf 62}(2), 467--475.
\item Morton, D.E., Saah, A.J., {\it et al.}, 1982. Lead absorption in children of employees in a lead related industry. {\sl American Journal of Epidemiology.} {\bf 115}(4), 549--555.
\item Neyman, J, 1923 [1990]. On the Application of Probability Theory to Agricultural Experiments. Essay on Principles. Section 9. {\sl Statistical Science.} {\bf 5}(4), 465--472. Trans. Dorota M. Dabrowska and Terence P. Speed.
\item Pruzek, R.M., Helmreich, J.E., 2009. Enhancing dependent sample analysis with graphics. {\sl Journal of Statistics Education.} {\bf 17} (1).
\item  Robins, J.M., Rotnitzky, A., Zhao, L.P., 1994. Estimation of regression coefficients when some regressors are not always observed. {\sl Journal of the American Statistical Association.} {\bf 89}(427), 846-866.
\item Robins, J.M., Rotnitzky, A., 1995. Semiparametric efficiency in multivariate regression models with missing data. {\sl Journal of the American Statistical Association.} {\bf 90}, 122--129.
\item Rubin, D.B., 1974. Estimating causal effects of treatments in randomized and nonrandomized studies. {\sl Journal of Educational Psychology.} {\bf 66}(1), 688-701.
\item Sakaguchi, S., 2016. Estimation of time-varying average treatment effects using panel data when unobserved fixed effects affect potential outcomes differently. {\sl Economics Letters.} {\bf 146}, 82--84.
\item Walker, S., 1996. An EM algorithm for nonlinear random effects models. { \sl Biometrics.} {\bf 52}, 934--944.
\end{description}

\end{document}